\documentclass{appolbprime}
\usepackage{epsfig}
\newcommand{\lsim}{\raisebox{-0.13cm}{~\shortstack{$<$ \\[-0.07cm] $\sim$}}~}
\newcommand{\gsim}{\raisebox{-0.13cm}{~\shortstack{$>$ \\[-0.07cm] $\sim$}}~}

\begin{document}
\title{Very Light Gravitino Dark Matter%
\thanks{Based on work in collaboration with Karsten JEDAMZIK (LPTA-Montpellier)
and Martin LEMOINE (IAP-Paris) }%
~\thanks{Presented at {\sl Physics at LHC, Cracow, Poland, 3-8 July `06} }
}
\author{Gilbert MOULTAKA 
\address{Laboratoire de Physique Th\'eorique et Astroparticules \\
{\sl UMR5207--CNRS}, Universit\'e Montpellier II \\
Place E. Bataillon, F--34095 Montpellier Cedex 5, France}
}
\maketitle
\vspace{-7cm}
\begin{flushright}
LPTA/06--57
\end{flushright}

\vspace{5.5cm}
\begin{abstract}
We address the question of dark matter in the context of gauge mediated
supersymmetry breaking models. In contrast with mSUGRA scenarios, the messenger
of the susy breaking to the visible sector can play an important role allowing
a relic gravitino in the $\sim \mbox{keV}$ to $10 \mbox{MeV}$ mass range
to account for the cold dark matter in the Universe.  
\end{abstract}
  
\section{Introduction: Neutralino versus Gravitino dark matter}

\noindent
It is certainly very attractive  that two longstanding open questions -the 
origin of electroweak symmetry breaking -and the nature of the non-baryonic 
dark matter in the Universe- seem to be on the verge of being answered
simultaneously and presumably in a unified framework. LHC will give us the 
opportunity to start scratching the surface of this issue and, if we are lucky 
enough, to hint more clearly at the correct unified framework. Direct and 
indirect searches for dark matter will also bring in a very interesting 
complementarity with the LHC and Tevatron searches. On the theoretical side, 
several avenues for physics beyond the Standard Model offer particle candidates 
for non baryonic dark matter. Among these candidates, the  (lightest) 
neutralino in the minimal supergravity (mSUGRA) scenario has been  so 
extensively studied that it deserves the status of  ``benchmark scenario'' 
\cite{king}. It should be clear, though, that this scenario is just a 
possibility among other equally compelling ones. Actually, one of its main 
advantages is its relative model-independence regarding early Universe issues, 
making such a scenario ``simpler'' to study (not more ``natural''!) for that 
matter. Let us recall briefly 
these early Universe issues, as they will be important for the subsequent 
discussion: {\sl {\bf i)}} the particle content of the Universe at the end of 
inflation is assumed to be described by the MSSM plus the graviton and the 
gravitino. That is, the hidden sector responsible for supersymmetry breaking 
[and/or its communication to the MSSM] is essentially heavier than the reheat 
temperature, is not produced early on and has no bearing on the later evolution
of the Universe. {\bf ii)} all the MSSM particles are initially 
in thermal equilibrium. {\bf iii)} the gravitino may or may not be in thermal
equilibrium, and in the former case its number density  depends strongly on 
the reheat temperature.
Point {\bf i)} is valid typically in gravity mediated 
susy breaking models of which mSUGRA is an example, where the hidden sector 
lies somewhere between the GUT and the Planck scales. Point {\bf ii)} is a
simplifying working assumption [which cannot be addressed further without 
a more concrete model for the production of the light particles (inflaton
couplings to these particles, decay, etc...).] Points {\bf  i)}, {\bf ii)}
validate in the mSUGRA-neutralino-LSP (and similar) scenarios a routine relic 
density calculation for {\sl thermally produced} dark matter. The 'naturalness'
of the scenario is then inherited from the fact that any thermally produced 
weakly interacting stable particle having a weak scale mass gives in general 
the right order of magnitude of the relic density. Then detailed calculations
delineate the regions of the parameter space consistent with WMAP as well
as with particle physics constraints, yielding a plausible answer to the dark
matter problem. However, point {\bf iii)} which typically leads to a 
gravitino problem remains completely non-tackled in this class of scenarios. 
This is so even in variations of the above scenario where a very
heavy gravitino is supposed to produce the bulk of the dark matter  
non-thermally through its decay, 
(such as in the anomaly mediated susy breaking 
scenario), or when the gravitino is the lightest susy particle (LSP)  but is
produced dominantly non-thermally through the decay of the next to lightest 
susy particle (NLSP). In particular, in the latter scenario one needs to assume 
arbitrarily a sufficiently low reheat temperature to keep the thermal 
production subleading! \\
Having all this in mind, there is yet another important question for the high
energy colliders, namely unraveling the origin of supersymmetry breaking. 
It is then quite natural to ask what happens to the dark matter issue if the
gravitational interaction plays only a minor role in this 
breaking (and its mediation to the MSSM), thus invalidating the above scenarios.  
We will address this question
hereafter in the context of a representative class of gauge mediated susy
breaking (GMSB) models {\cite{GMSB, GR99}) which can be probed at the 
LHC. Some features of points {\bf i)} to {\bf iii)} are modified in this 
context and we will argue that these modifications can allow for a very light
 (but still cold) gravitino dark matter freed in the same time from a gravitino 
problem. 

\section{Supersymmetry breaking through gauge mediation}

\noindent
If supersymmetry is realized in nature, not only would this give us confidence
in our understanding of the large hierarchy stabilization
 between the electroweak scale $G_F^{-1/2}$ and the  GUT or Planck scales,
but also the hope that its dynamical breaking 
would 'explain' this hierarchy: one would expect typically  
$G_F^{-1/2} \sim {<F> \over M }$ where $M$ is the mass scale
of some supersymmetric hidden sector and ${<F>^{1/2}}$ the mass scale
of its corresponding susy breaking communicated to the MSSM. Another  
relation coming from the supergravity sector and entailing a very small
cosmological constant relates the gravitino mass $m_{3/2}$ to 
the total susy breaking mass scale ${<F_{tot}>^{1/2}}$ through  
$m_{3/2} \simeq {<F_{tot}>}/({\sqrt{3} m_{\rm Pl}} )$ , where $m_{\rm Pl}$ is the 
(reduced) Planck mass and $<F> \;  \leq  \; <F_{tot}>$. The resulting
relation $G_F^{-1/2} \lsim m_{3/2} \left({m_{\rm Pl} \over M} \right) $
implies qualitatively that in gravity mediated susy breaking models 
($M \simeq m_{\rm Pl}$), $m_{3/2}$ is of order 
the electroweak scale,  while it becomes much smaller when the susy breaking 
is essentially mediated by a gauge sector ($M \ll m_{\rm Pl}$), the
gravitino thus becoming the LSP. In the present study 
we will be interested in the range 
$m_{3/2} \sim {\cal O}(1) \mbox{keV} - {\cal O}(1) \mbox{GeV}$. Moreover, the 
considered GMSB models \cite{GMSB} have two separate sectors on top of the MSSM: 
a secluded sector where the dynamical susy breaking takes place with no
direct couplings to the MSSM, and a messenger sector charged under the
standard model gauge groups thus having gauge interactions with the MSSM
particles. A spurion field couples directly 
to the messenger sector transferring to it part of the susy breaking effects,
the latter being ultimately carried further to the MSSM via the gauge couplings 
of the messengers. In particular the gaugino and scalar soft susy breaking 
masses of the MSSM, $m_{1/2}^i$, $m_0^s$ are generated respectively to one 
and two loop orders in the form  
$\sim \left({\alpha_i\over 4\pi}\right){<F_S>\over M_X}, 
\sqrt{\left({\alpha_i\over 4\pi}\right)^2 \kappa^i_s}{<F_S>\over M_X}$. Here $i$ 
labels the three gauge couplings of the standard model, $s$ the scalar quarks
and leptons and Higgses, $\kappa^i_s$ a numerical factor depending on the
messenger number and representations, $<F_S>$ the partial
susy breaking contribution transmitted by the spurion $S$ to the visible
sector, and $M_X$ is the mass scale of the messenger sector. The ensuing 
universality of $m_0^s$ for each flavour sector at the messenger scale 
guarantees the absence of flavour changing neutral currents (FCNC).      
Furthermore,  the low susy breaking scale in GMSB models may be 
favoured by the issue of the little fine-tuning problem \cite{falkowski}.

\section{Gravitino problem - Messenger solution}

\noindent
Depending on the value of the reheat temperature $T_{RH}$ subsequent to an 
initial inflationary phase of the Universe, the secluded and messenger sectors 
of GMSB may or may not be produced in the early Universe. As we stressed in 
the introduction this is in contrast with mSUGRA and leads to a modification of 
assumption {\bf i)}. Though the production of these sectors is {\sl \`a priori}
a complication it can also be a blessing. We will illustrate this aspect 
hereafter by considering the spurion and the messenger fields. The mass
degeneracy within a supermultiplet of messenger fields is lifted by susy
breaking leading to a lighter and a heavier scalar messengers with masses
$M_\pm =M_X(1 \pm \frac{<F_S>}{M_X^2})^{1/2}$ and a fermionic partner with mass 
$M_X$. Thus $\frac{<F_S>}{M_X^2} < 1$. 
Moreover, one has to require $\frac{<F_S>}{M_X} \lsim 10^5 \mbox{GeV}$ to 
ensure an MSSM spectrum $\lsim {\cal O}(1) \mbox{TeV}$. One
 then expects typically $M_X \gsim 10^5 \mbox{GeV}$.
On the other hand taking for example a gravitino mass 
$m_{3/2} \simeq 1 \mbox{MeV}$ would imply typically an
upper bound $(T_{RH} \lsim )10^5 \mbox{GeV}$ on the reheat temperature above
which the thermally produced gravitino [via scattering of strongly interacting 
MSSM particles with a gluino mass $\sim 1 \mbox{TeV}$] 
would overclose the Universe \cite{TMY93}. This particular configuration illustrates
qualitatively the possible interplay between the gravitino and the messengers: 
if $T_{RH} \lsim M_X$ only the MSSM is present and simultaneously the gravitino 
relic density is acceptably small. If  $T_{RH} \gsim M_X$ we have a gravitino
problem, but since now the messenger sector is also (partly or wholly) produced
it will on one hand contribute to the thermal gravitino production through
scattering processes, and on the other hand will pause a cosmological problem 
on its own! Indeed in typical GMSB models the lightest messenger particle (LMP)
with mass $M_{-}$ is stable due to the conservation of a messenger quantum 
number. Its thermal relic density is calculable similarly to that of the mSUGRA
neutralino LSP and is found to scale as 
$\Omega_M h^2  \simeq 10^5 \left(\frac{M_{-}}{10^3 TeV}\right)^2$, thus
overclosing the Universe in most of the parameter space under consideration.
A straightforward solution to this problem is to let the LMP decay into MSSM
particles. An interesting scenario was proposed \cite{FY02} where this LMP
decay leads to a substantial increase of entropy  thus solving {\sl also} 
the gravitino problem by diluting its relic density to a level which can 
account for the dark matter in the Universe. For this scenario to work, though, 
a few necessary conditions are required which delineate the favourable parts 
of the parameter space: for instance, the LMP should dominate the Universe 
energy density before it decays, but it should decay after gravitino has 
freezed-out from the thermal bath. A typical configuration $T_d < T_{MD} <
T^f_{3/2}$ where $T_d, T_{DM}, T^f_{3/2}$ denote respectively the LMP decay
and matter domination temperatures, and the gravitino freeze-out temperature, 
is determined by the particle properties (annihilation cross-section and
decay width of the LMP, etc...). The entropy release, diluting the initial
gravitino density, is determined by 
the temperatures before and after LMP decay. But the final gravitino relic
density can also receive substantial thermal and/or non-thermal contributions 
from LMP scattering or decay, depending on the detailed assumptions of the 
model which we briefly describe in the following sections.

\section{coupling to supergravity \& GUT groups}

\noindent
Although (super)gravity plays no role in breaking supersymmetry in GMSB models,
there are still a few reasons for considering its full coupling to the model: 
-the gravitational sector provides a natural framework
for an unstable LMP -the complete couplings of the gravitino (and the graviton!)
to the MSSM as well as to the spurion and messenger sectors are needed for a 
reliable estimate of the cosmological constraints, (not to mention the phenomenological
need to reabsorb the goldstino degrees of freedom in the massive gravitino.)
In the gauge sector the stability of the LMP is usually achieved by a 
discrete symmetry conserving the messenger number. Breaking explicitly 
this symmetry at low scales would ruin the natural suppression of FCNC in 
GMSB models (a crux of these models), unless the new couplings are unnaturally 
suppressed.  In contrast, Planck scale physics arguably breaks
discrete symmetries (at least when they are not residuals of broken continuous
symmetries). Messenger number non-conserving effective operators  
are then expected, which will in most cases fall into two classes leading to 
slow decays of the LMP into MSSM particles with a suppression  ${\cal O}(m^2_{3/2})$ 
or  ${\cal O}(m_{\rm Pl}^{-2})$ \cite{JLM04}. The proposal in \cite{FY02}  
belongs to the first one of these two classes, however, as shown in 
\cite{JLM04} and illustrated in the next section, taking into account all the 
supergravity effects leads to modified results. 

In order to preserve gauge coupling unification it is sufficient to assume
that the messenger fields sit in complete GUT group representations \cite{GR99}.
 In \cite{JLM04, JLM05} we have studied somewhat in detail 
the impact of the usual assignments ( $\mathbf{5}_M+\mathbf{\overline{5}}_M$ or
$\mathbf{10}_M+\mathbf{\overline{10}}_M$ of $SU(5)$ and 
$\mathbf{16}_M+\mathbf{\overline{16}}_M$ of $SO(10)$) on the gravitino DM
scenario. A qualitative difference between $SU(5)$ and $SO(10)$  is that in
the latter the LMP is an MSSM singlet. Its annihilation to standard model
particles is one-loop suppressed \cite{JLM05}, leading typically to much larger
LMP relic density than in the $SU(5)$ case for comparable LMP masses, whence a
larger entropy production due to its decay and a smaller gravitino relic
density.

\section{Gravitino relic density}

Let us first list all the ingredients entering  the
gravitino relic density calculation:

\noindent
$\bullet$ the MSSM, the spurion and the messenger sector are all produced
at the end of inflation (i.e. sufficiently high $T_{RH}$).

\noindent
$\bullet$ the gravitino relic density  breaks up into 
$\Omega_{3/2}= (\Omega_{\rm th.} + \Omega_{\rm non-th.}) \!\times \! \Delta^{-1}$
where $\Omega_{\rm th.}= \Omega_{\rm scatt.} + \Omega_{\rm dec.}$ is the 
contribution from scattering and/or decays of particles in the thermal
bath, $\Omega_{\rm non-th.}$ is the contribution of (slowly) decaying particles
such as NLSP or LMP into gravitinos 
{\sl subsequent} to the decoupling of the latter from the thermal bath, 
and $\Delta (\gg 1)$ is a dilution factor due to the late decay of the LMP
into MSSM particles as discussed in section {\bf 3}.  


\noindent
$\bullet$ the decay of the LMP is induced by Planck scale messenger number 
non-conserving operators present either in the K\"ahler potential or in the 
superpotential. An exhaustive study is carried out in \cite{JLM04}.
Here we consider for illustration two such non-minimal contributions to the
K\"ahler potential,  
$\delta K_1 = {\mathbf 5}_M\overline{\mathbf 5}_F + h.c.$ and 
$\delta K_2= \mathbf{5}_M\mathbf{\overline 5}_{F}\mathbf{24}_H/m_{\rm Pl} + 
h.c.$, where $\mathbf{5}_M, \mathbf{5}_F, \mathbf{24}_H$ are superfield
$SU(5)$ multiplets respectively of the messengers, the MSSM matter fields and
the GUT Higgs fields.   

\noindent
$\bullet$ an important characteristic of $\delta K_1$ ($\delta K_2$) is that 
it induces, after supersymmetry (and GUT symmetry) breaking, 
$m_{3/2}$ ($m_{3/2} M_{\rm GUT}/m_{\rm Pl}$) suppressed effective couplings
leading to LMP two-body decays into a lepton and a gaugino, a Higgs and a 
slepton, a Higgsino and a matter fermion, or three-body decays into
a sneutrino and two gravitinos. Some of these decays can affect the light
elements abundance as predicted by the standard Big-Bang Nucleosynthesis (BBN),
others can inject a hot/warm gravitino dark matter component leading to
interesting constraints.

\noindent
$\bullet$ finally one has also to consider the effect of the gravitationally 
induced LMP annihilation into a pair of gravitinos which involves graviton
and spurion exchanges and depends on whether the spurion is much heavier
or lighter than the LMP. 

In figures 1 and 2 we show the results  respectively for $\delta K_1$ and
$\delta K_2$ operators, in the plane $m_{3/2} - M_X$  (the messenger mass
scale) and assuming a reheat temperature $T_{RH}= 10^{12}$ GeV, a $150$ GeV
neutralino NLSP and a $1$ TeV gluino.   Left (right) panels correspond to a
spurion heavier (lighter) than the LMP, leading to quite different
behaviour of the LMP annihilation into gravitinos as shown in the figures. The
horizontally red-line shaded  areas in the left upper corners are physically 
excluded
[they would correspond to a total susy breaking smaller than the fraction
communicated to the visible sector!]. The fully colored (blue, red, green and
yellow) areas indicate the regions where the gravitino relic density is
cosmologically acceptable; the white area is where the gravitino overcloses the
Universe. Note that without the GMSB messenger sector all the $m_{3/2}$ range
would have been excluded \cite{TMY93} given the value of $T_{RH}$. The yellow
area in the r.h. panels corresponds to $\Omega_{3/2} < 0.01$ thus
solving the gravitino problem but not the DM issue. The blue, red and green
areas  correspond respectively to hot, warm and cold  gravitinos accounting for 
the dark matter. As can be seen in the r.h. panels of the two figures, cold 
gravitino dark matter occurs for 
$10\mbox{keV} \lsim m_{3/2} \lsim 1 - 10  \mbox{MeV}$ and a messenger scale  
$10^8\mbox{GeV} \lsim M_X \lsim 10^{11} \mbox{GeV}$. It is also interesting to 
note, in relation to structure formation issues, the possibility of mixed  
warm/cold DM in this range of parameters as can be seen in Fig. 1. 

\begin{figure}[h]
\caption[...]{The LMP decays through $\delta K_1$}
\includegraphics[width=1.\textwidth, height=1. \textheight, 
keepaspectratio]{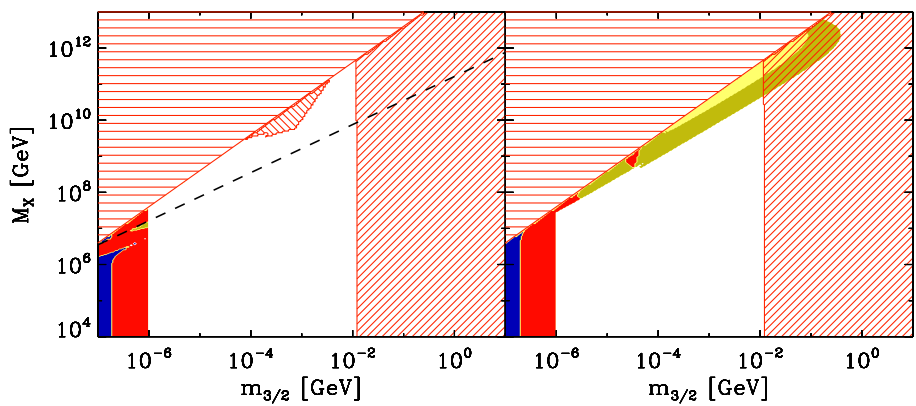}
\vspace{-.5cm}
\label{fig1}
\end{figure}

\vspace{-.3cm}
\begin{figure}[h]
\vspace{-.5cm}
\caption{The LMP decays through $\delta K_2$}
\includegraphics[width=1.\textwidth, height=1. \textheight, 
keepaspectratio]{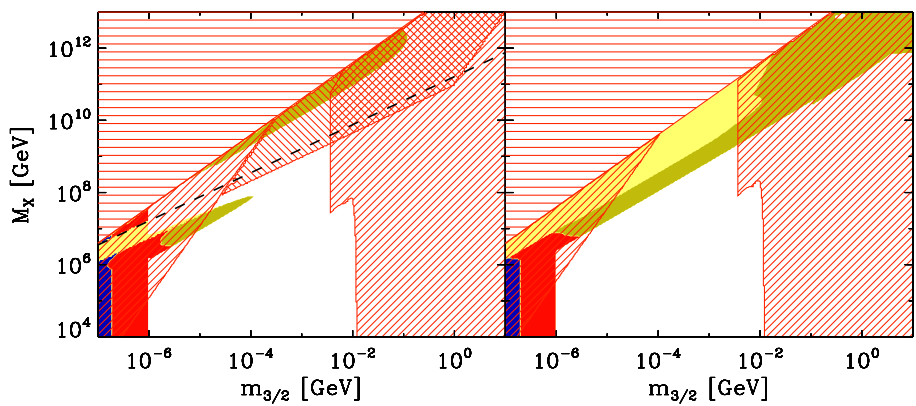} 
\end{figure}

\vspace{-.5cm}
\noindent
In the left-hand panels  where the spurion is much heavier than the
LMP, theoretical uncertainties occur above the black dashed line signaling
a saturation of unitarity through multi-gravitino production. Moreover
a significant part of the potential DM solutions is excluded in this case
by BBN constraints [e.g. in Fig.2, the areas shaded by NE-SW oriented red lines 
denote too slow NLSP (LMP) decays for large (small) $m_{3/2}$ and the 
NW-SE red-line shaded area corresponds to too energetic gravitinos].

\section{signatures at the LHC}

\noindent
The coupling of a light gravitino to matter scales like $<F_{tot}>^{-1}$
through the goldstino component. 
This makes the detection in direct and indirect dark matter searches 
deceptively hopeless in the mass range under consideration. The colliders 
become then a unique place to look for
distinctive signatures. A favourable situation for the LHC would be a charged
unstable slepton NLSP  with  $.5 \mbox{m} \lsim c\tau \lsim 1 \mbox{km}$ 
\cite{AMPPR01}, e.g. $c \tau \simeq 50 -200 \mbox{m}$ 
 for the lower part of the DM $m_{3/2}$ range considered in the previous
 section, taking $m_{NLSP} =150 - 200$ GeV. In the upper part of the allowed
 $m_{3/2}$ range [which can be even larger than in Figs.1, 2, due to less
 restrictive BBN bounds] the slepton decays typically outside the detector but 
 still yields a distinctive charged track. A neutralino NLSP would be a more
 difficult scenario with a $c \tau \gg {\cal O}(1) \mbox{m}$, \cite{KKNO03}.
 Distinguishing this case from a truly LSP neutralino would require indirect
 and more model-dependent information from other sectors of the MSSM
 (reconstruction of the would be neutralino relic density, signals from dark 
 matter searches,...). 


{\sl Acknowledgements: I would like to thank the organizers for the kind 
invitation and for the warm and friendly atmosphere during the conference.} 

\end{document}